# 10 Inventions on Improving Toolbars
## A TRIZ based analysis

**Umakant Mishra**

Bangalore, India
http://umakantm.blogspot.in

**Contents**

# 1. Introduction

Toolbar is one among the popular elements of a graphical user interface. The other popular elements of graphical user interface are buttons, menu, scrollbar, dialog box etc., all of which provide easy access to various functions of a GUI system.

A toolbar often does a similar function as the menu but with certain differences. A menu has the advantage of holding a large number of items without needing any additional screen space. In contrast, each button on the toolbar permanently occupies some space on the screen. It's not possible to implement large number of functions through a toolbar, as they will occupy more and more valuable screen space. However, the toolbar has an advantage as it gives a single click access to any function unlike a menu system where the user has to navigate through sub-menus to ultimate discover the item he is looking for.

## 1.1 Desired features of a toolbar

- A toolbar consists of several icons each of which is linked to a function. The user is supposed to press the icons through a pointing device like mouse to invoke the linked function.

- The toolbar may have either textual captions or graphical icons or both. Many current day toolbars consist of only icons.

- There may be multiple toolbars displayed on a single window. The user may selectively show and hide toolbars.

- The icons in the toolbar should be meaningful to explain their functions, e.g., a printer icon represents printing of a document.

- The buttons may have two or more states, e.g., elevated and depressed. The different states are represented in different brightness of color.

- Non-available icons may be shown in a different color.

- Moving the pointer on to a toolbar icon may show some hint regarding the function associated with that icon.

- There may be button groups, dropdown lists and other types of components in the toolbar.

- The user should be able to configure the icons on the toolbar intuitively. The configuration mechanism should be easy.

- The toolbar should be adaptive. In other words, the system should intuitively configure the icons on the toolbar based on various factors.

## 1.2 Ideal features of an advanced toolbar

- The toolbar should be adaptive to show and hide icons as may be required by the user from time to time.

- The toolbar should automatically rearrange the icons wherever necessary for easy finding of frequently used tools.

- The toolbar should not occupy too much space on the screen. In other words the toolbar should use minimum amount of screen space.

- Scrolling and dropdown features to accommodate more items.

- The icons on the toolbar should be self explanatory or easy to recognize.

- Resizing the fonts and icons based on user visibility and refreshing the look and feel based on user choice.

## 1.3 Problems in achieving the ideal features

When more and more items are added to the toolbar the size of the toolbar increases which occupies valuable screen space. This effectively reduces the size of the screen available for displaying valuable data or image. This situation leads to a contradiction. "The toolbar should have more items to provide more functionality, but it should have fewer items to occupy less screen space."

The other problem is flexibility and adaptability of the toolbar. The developer adds all the functions to a toolbar, as he does not know the exact requirements of a user at a future period of time. The user knows the exact requirements but he cannot change a fixed toolbar compiled inside the program. This situation leads to a contradiction. "The program should be compiled so that the source code is not altered by others, but the user should be able to alter and reconfigure the toolbar items."

## 1.4 Applying Inventive Principles in solving toolbar problem

**Problem**: When the number of controls in the toolbar is more, the size of the toolbar may not fit into the size of the window. In that case part of the toolbar may not be visible to the user thereby making some important controls inaccessible.

**Solution**: Isolate the toolbar from the application window (Principle-2: Taking out) and use a global toolbar (Principle-6: Universality). A global toolbar is typically independent from an application window and is displayed on the top or side of the screen thus getting more space.

**Limitation**: However the toolbar can be maximum as long as the width (or height in case of vertical toolbar) of the monitor screen. Besides, the universal toolbar suffers from other drawbacks.

**Alternative Solution**: Keep the labels at the right side of the icon since the width of the screen is more than its height (Principle-17: Another dimension).

**Limitation**: still the number of icons are limited to the space availability.

**Solution**: Remove the labels and display only the icons (Principle-2: Taking out).

**Limitation**: the advantage of having labels is lost.

**Solution**: Display the labels or hints when the pointer is on top of the icon. This will give the advantage of the labels without occupying screen space (Principle-17: Another dimension).

**Limitation**: Still the number of icons are limited to the available space.

**Solution**: Display only the most important icons based on a priority. In other words, display the icons and labels, which are deemed most important on the available toolbar space (Principle-16: partial or excessive action).

**Limitation**: The most important icons are calculated for the father (one user) who was using the system, but the son (another user) working on the same computer may not find those icons useful.

**Solution**: Use multiple heuristic criteria to prioritize the icons (Principle-40: Composite). Maintain separate priority lists or preferences for individual users (Principle-3: Local quality).

**Limitation**: What if the window size is so reduced that even the most important (or most frequently used) icons cannot be displayed.

**Solution**: If the space is more then add more icons and if the space is less then remove low priority icons (Principle-15: Dynamize).

# 2. Inventions on improving Toolbars in GUI

There are several aspects of a toolbar such as its functionality, flexibility, size, orientation, and aesthetics etc., which contribute to the overall effectiveness of it. There are several inventions on improving various aspects of a toolbar. The following are ten interesting cases selected from US patent database.

## 2.1 Combination of both static and dynamic tool palette

**Background problem**

Some windows applications provide a static tool display. In a static tool display, all tools remain displayed regardless of their relevance to the current module of the application. This method is confusing as the operator may attempt to select a tool, which is not selectable.
Some other windows applications provide a dynamic tool palette display in which tools are added and subtracted as they become applicable to the current module of the application. This method is also confusing as the commonly used tools are shifted around in the display.

**Solution provided by the invention**

Patent 5572648 (Invented by Bibayan, Assigned to Canon Kabushiki, Nov 96) discloses a method of simultaneously displaying a static tool palette having predefined windowing tools and a dynamic tool palette that changes the windowing tool functions in accordance with the context of the executed application program.

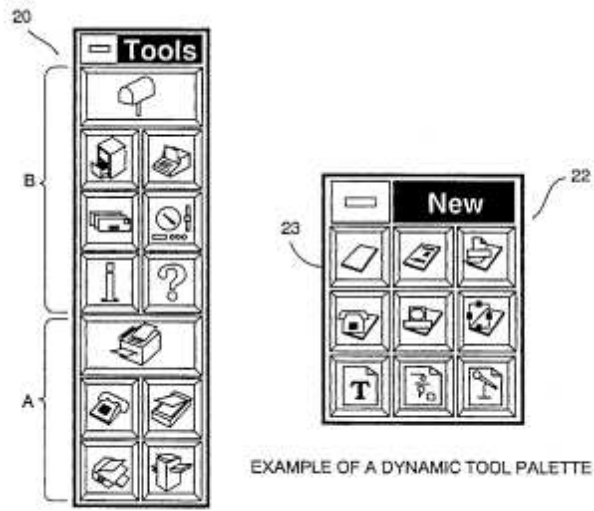

EXAMPLE OF A DYNAMIC TOOL PALETTE

**TRIZ based analysis**

The invention displays both a static tool palette and a dynamic tool palette to get the advantage of both the methods (Principle-5: Merging).

## 2.2 Method and system for stacking toolbars in a computer display

### Background problem
As the space of a GUI is limited, there is a limit to the number of user interface elements that can be displayed to a user. When a toolbar is selected by the user the space occupied by the toolbar becomes unavailable for display of other information or data. This forces the user to compromise on either closing the toolbar or loosing to view the valuable data.

### Solution provided by the invention
Patent 5644737 (invented by Tuniman et al., Microsoft Corporation, July 1997) disclosed a stacking arrangement of toolbars. Only the graphic objects on one or more selected toolbars are displayed. The user can selectively choose a toolbar either to display or hide. The selected toolbar becomes fully visible and covers the non-selected toolbars. When a toolbar opens or closes it slides to different positions with animation and an audible sound for added realism.

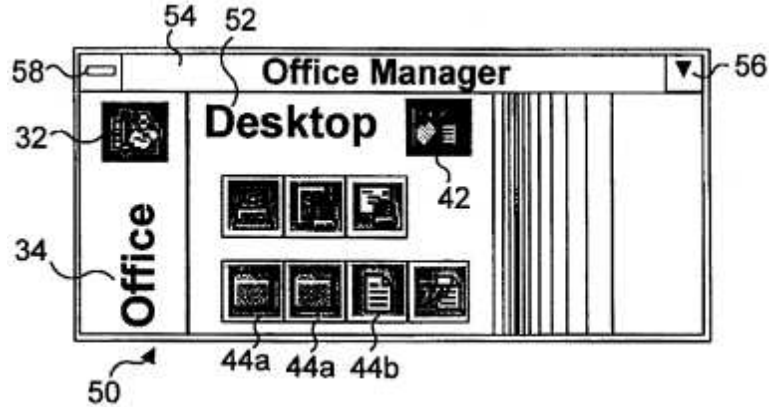

### TTRIZ based analysis
The invention uses stacking of multiple toolbars on the same area to preserve screen space (Principle-7: Nested doll).

The user can selectively open and close the toolbars (Principle-15: Dynamize).

While opening or closing the toolbar, it slides in and slides out with animation and sound to produce enhanced realism (Principle-38: Enrich).

## 2.3 Method and system for adding buttons to a toolbar

### Background
The toolbars provide convenient alternatives to drop down menus. Some of the modern toolbars provide options to add and remove buttons from the toolbar, but through difficult and cumbersome methods. There is a need to provide an easy way of adding and removing buttons from a toolbar.

**Solution provided by the invention**

US Patent 5644739 (Invented by Elizabeth Moursund, Assigned to Microsoft Corporation, Jul 97) disclosed a method of intuitively adding a button or other type of control to a toolbar. According to the invention dragging and dropping controls onto the toolbar can create new toolbar controls. The new control is bound to an operation of the object and can be executed by a mouse event.

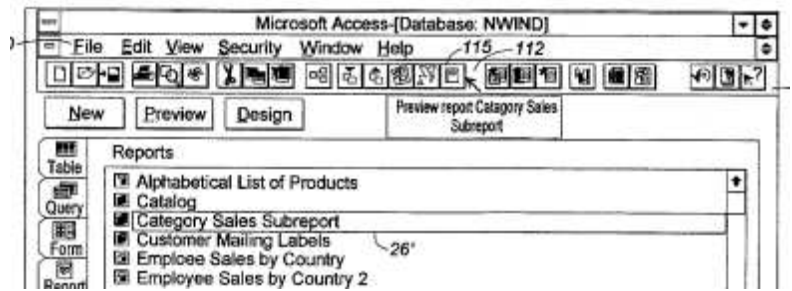

**TRIZ based analysis**

The invention allows dragging and dropping items on the toolbar to create buttons for the links. This method facilitates easy adding and removing buttons to the toolbar (Principle-15: Dynamize).

**2.4 Sliding out interface**

**Background problem**

The recent application programs are shipped with hundreds of command buttons, which are very difficult to display within the limited screen space. There are several solutions to accommodate more icons in the limited screen space. One is by reducing the icon size, but this solution still occupies permanent screen. Another method is by grouping the related commands into folder icons, but this method is difficult to explore by clicking on folder icons.

**Solution provided by the invention**

Rubin et al. disclosed a sliding out interface bar (Patent 5914716, assigned to Microsoft corporation, Jun 99), which conserves valuable screen space. The graphical user interface displays a target image representing a sliding out command bar containing a set of selectable computer resources. When the user moves the display pointer to a location near the target image, the tray slides out and the command bar remains visible. When the pointer is moved away from the image, the command bar slides in and becomes invisible.

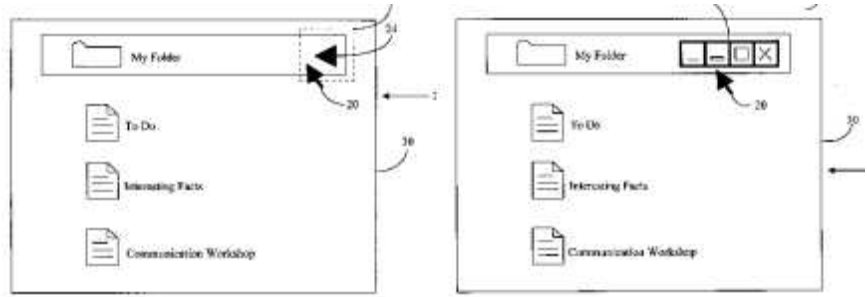

### TRIZ based analysis

We want hundreds of icons to be displayed on the screen, but we don't want them to occupy valuable screen space (Contradiction).

One conventional solution is to reduce the size of the icon, which will effectively occupy less screen space (Principle-35: Parameter change). Another solution is to group the related commands and keep inside folder icons (Principle-5: Merging).
This invention discloses a sliding interface, which slides out when the user moves the pointer near to it and slides in when the pointer is moved away (Principle-15: Dynamize).

### 2.5 System and method for resizing and rearranging a composite toolbar by direct manipulation

### Background problem

Some advanced toolbars are user configurable. They give user the option to add new buttons or delete existing buttons from the toolbar. The button manipulation is generally accomplished by interacting with a separate dialog box.

Secondly, when the toolbars are moved to various locations there is no mechanism for automatically filling in gaps between the toolbars. This requires the user to manipulate the displayed toolbars one by one to fill in the gaps between toolbars.

There is a need for a method that supports a direct manipulation of toolbars on the computer display screen itself. There is also a need for a method of operating multiple toolbars simultaneously.

### Solution provided by the invention

US Patent 6057836 (invented by Kavalam et al., Assigned to Microsoft Corporation, May 2000) Discloses a method of resizing and rearranging a toolbar directly on the screen. As per the invention the composite toolbar is displayed in an initial configuration state with resizing areas. The user resizes the toolbar by dragging the resizing area.

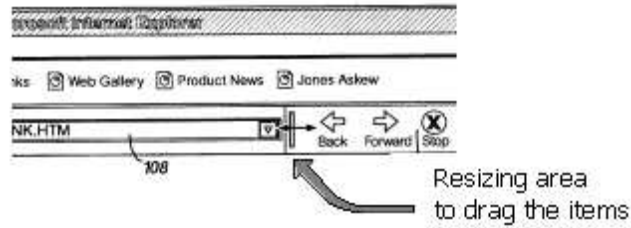

According to the invention the user can drag a section expand or collapse that section. When the section is expanded, it overlaps an adjacent section. Similarly, when the section is collapsed it reveals the adjacent section.

**TRIZ based analysis**

The invention intends to manipulate the buttons and adjust the size of the toolbar (Principle-15: Dynamize).

The invention avoids a configuration dialog box and allows direct manipulation of the toolbar on the screen (Principle-21: Skipping).

The method utilizes sliding and overlapping features as a mechanism of customizing toolbars (Principle-15: Dynamize).

**2.6 Customizing toolbar through a quick customize menu**

**Background problem**

There are various methods to make a toolbar more flexible and customizable. Some systems provide turning on or off the display of any toolbar and customize the items in the toolbar. Some systems provide a drag and drop system to add or remove controls. But these customizations include sophisticated process. There is a need for a method to easily customize the controls in a toolbar.

**Solution provided by the invention**

US Patent 6133915 (invented by Aurcuri et al., Assigned to Microsoft, Oct 2000) invents a method for customizing the toolbar through a quick customize (QC) menu. The QC menu displays a list of controls associated with the toolbar for user to select and unselect controls to be displayed on the toolbar.

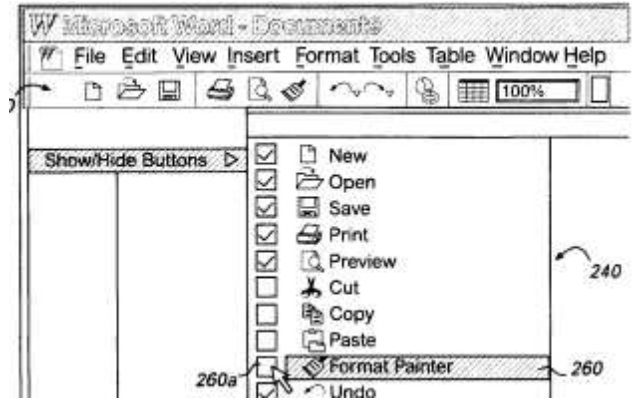

**TRIZ based analysis**

The invention provides a quick customize menu ("QC menu") for selecting the controls on the toolbar. (Principle-24: Intermediary).

By enabling a selection indicator, the corresponding control is presented on the toolbar, whereas by disabling the selection indicator removes the corresponding control from the toolbar. (Principle-35: Change parameter).

**2.7 Displaying most important controls on the toolbar**

**Background problem**

The toolbar generally displays the icons and controls for a single click access to the most frequently used operations. Usually the developer has to determine the importance of the controls in anticipation with the users needs. Sometimes the developer has to do a guesswork on which controls are more or less important for the user. This method creates dilemma for the developer and restriction for the users.

It is necessary to display the controls on the toolbar based on the need and usage pattern of the user. In other words, the system should find a mechanism to display the most important controls on the toolbar, which may change throughout the session.

**Solution provided by the invention**

Patent 6232972 (Invented by Arcuri, et al., assigned to Microsoft, in May 01) provides a method of displaying the controls on the toolbar, which are most important to the user. The invention tracks the usage of the controls to determine which controls are more important and which controls are less important. It places the more important (or more likely to be used) components on the toolbar and removes the less important (or less likely to be used).

When the toolbar is loaded for the first time, it loads the controls based on a predefined order. Subsequently the priority changes according to the usage pattern of the controls.

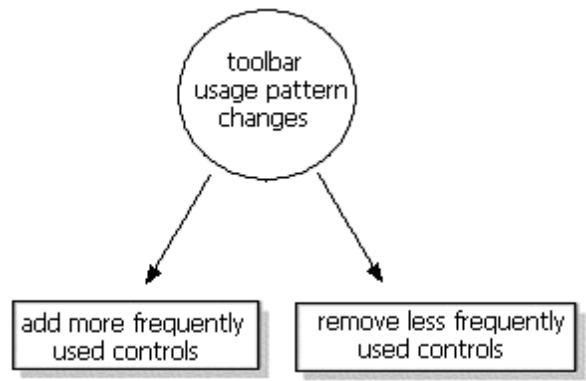

**TRIZ based analysis**

The invention divides to controls in the toolbar according to their "most recently used" status. The most recently used controls are displayed in the toolbar and others are displayed in a drop-off well (Principle-1: Segmentation).

When the user selects a control from the drop-off well display, the status of selected control is changed to "most recently used" (Principle-23: Feedback, Principle-35: Change parameter).

If the space in the toolbar does not allow displaying all the recently used controls, the number of controls to be displayed are calculated based on the "most recently used" controls and the size of the associated control. (Principle-16: Partial or excessive action).

**2.8 Procedural toolbar user interface**

**Background problem**

Most dropdown menus are static and follow a predefined command structure. This structure is not suitable for displaying a context sensitive menu that can display items based on the options previously selected by the user similar to the mechanism of a wizard. There is a need for an improved interface that can automatically present context sensitive options based on previous selections.

**Solution provided by the invention**

Patent 6456304 (Invented by Angiulo et al, Assigned to Microsoft, Sep 2002) provides a user interface toolbar that enables a user to make selections in a procedural order. The toolbar provides a plurality of controls each containing a control value and a dropdown menu. The menu of a given selection control depends on the value of the selection control that is to its immediate left so that the menus are nested on a context sensitive basis. Unlike a wizard, which does not allow re-entering previously selected options, this mechanism allows a user even to change the value of a previously selected control.

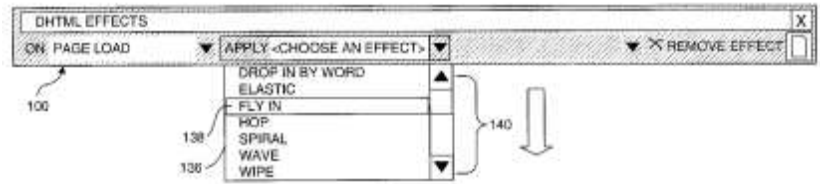

For example, the menu items in the first selection control depend on a context. The menu items in the second selection control dropdown menu are based on the item selected in the item selected in the first selection control. Similarly the menu items for the third selection control dropdown menu are context sensitive to the value selected in the second selection control and so on.

The toolbar preferably includes a "clear all" control to clear all control values, a "highlight selected component object" toggle button, and other features.

**TRIZ based analysis**

The invention dynamically changes the options in the menu depending on the previous selections made by the user (Principle-15: Dynamize, Principle-23: Feedback).

The user can even change the value of a previously selected control. In such a case, the downstream values are either updated or cleared depending on their validity in the context (Principle-15: Dynamize).

**2.9 Easy method of dragging pull-down menu items onto a toolbar**

**Background problem**

Accessing an option through navigating a menu tree is time taking. This is worse in case of a sub-menu item which needs to activate and go through several level of menu by controlling the mouse pointer. Selecting an item on a toolbar is faster as it does not require activation of any menu. On the other hand a toolbar permanently occupies some real estate on the GUI.

**Solution provided by the invention**

US Patent 6621532 (invented by Mandt, assigned to IBM, Sep 03) provides a method of dragging pull down menu items onto a toolbar. According to the invention when the user drags a menu item and drops on the toolbar, the menu is automatically converted to a toolbar button. This facilitates the user to easily access the option during later use.

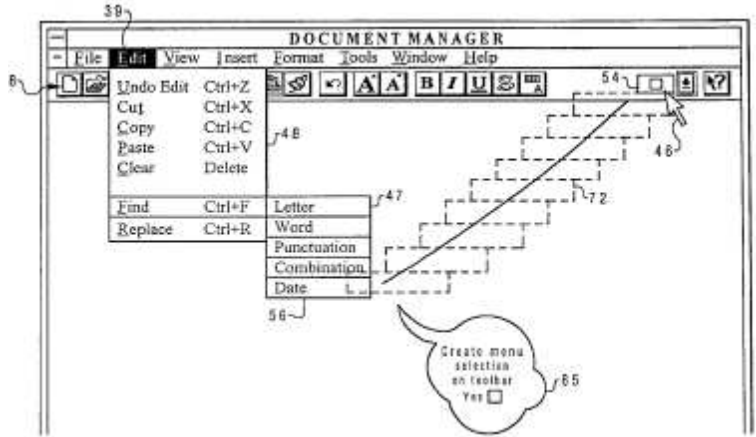

**TRIZ based analysis**

The menu mechanism consumes less screen space but navigation through a menu is difficult, as the user has to click several times. If menu items are displayed permanently like toolbar items, they will occupy more screen space (Contradiction).

The invention provides option to drag more frequently used option from a sub-menu to the toolbar. This method makes the item accessible by a single click (Principle-35: Parameter change).

**2.10 Dynamically adjustable toolbar**

**Background problem**
When a toolbar is visually attached to a particular application window, its length will be restricted to a particular application window. There restricts on the length of the toolbar and limits the number of buttons that can be included in the toolbar. The icons and labels cannot be too small as they should be recognizable by the user. Again application running on lower resolution can still contain less number of icons.

**Solution provided by the invention**
Patent 6624831 (Invented by Shahine, et al., assigned to Microsoft, Sept 2003) invents a method of effectively dealing with the size limitation and changes in size of the toolbar. The invention discloses a dynamically adjustable toolbar where the labels and icons are added or removed based on an assigned priority.

When the toolbar is first displayed it considers all the items associated with it. The initial priority is decided by a pre assigned weightage. But the priorities of the items are gradually changed according to the frequency of their usage. Then depending on the space availability the buttons are selected based on their assigned priority. The button icons and labels having higher assigned priorities are displayed before those having lower priorities.

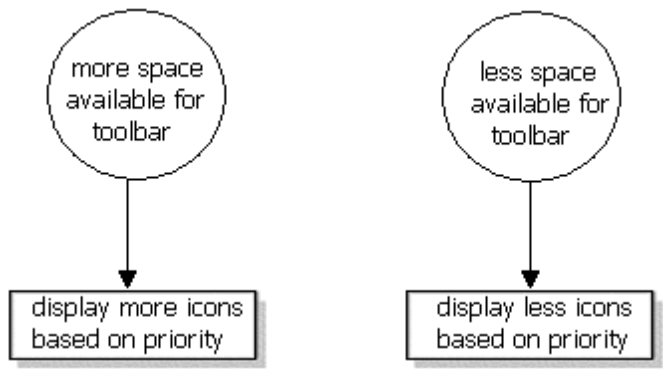

**TRIZ based solution**

The invention displays only the most important icons based on a priority (principle-16: Partial or excessive action).

The invention adds more icons when there is more space and removes additional icons when there is less space (Principle-15: Dynamize).

## 3. Summary and conclusion

The major objectives of a effective toolbar is to:

- Offer maximum buttons in minimum screen space.
- Easy configuration or auto configuration of toolbar icons.
- Easy recognition and easy navigation of the toolbar icons.

An advanced toolbar is very different from a conventional static toolbar. The above inventions have tried various ways to enhance the capability of a toolbar. Some interesting enhancements are:

- Auto adjustable features to display and hide icons according to usage pattern, adjusting size of the toolbar according to window size.
- Allowing user to configure icons, shape, size, color, position, look and feel and other aspects of the toolbar.
- Enhancing user friendliness by improving navigation, assistive features, aesthetics and other means.

The inventions illustrated above try to improve various aspects of a toolbar. As toolbar is seen as one of the essential components of a graphical user interface we can expect to see more and more inventions on a toolbar trying to improve its features while overcoming the current limitations.